\documentclass[sigconf]{acmart}

\RequirePackage[l2tabu, orthodox]{nag}

\usepackage[ruled,vlined,linesnumbered]{algorithm2e}
\usepackage[english]{babel}
\usepackage{booktabs}
\usepackage[T1]{fontenc}
\usepackage{graphicx}
\usepackage[htt]{hyphenat}
\usepackage[utf8]{inputenc}
\usepackage{listings}
\usepackage{url}
\usepackage{xcolor}

\definecolor{color-bg}{HTML}{F6F8FA}
\definecolor{color-keyword}{HTML}{D73A49}
\definecolor{color-ident}{HTML}{005CC5}
\definecolor{color-string}{HTML}{032F62}
\definecolor{color-comment}{HTML}{6A737D}

\copyrightyear{2021}
\acmYear{2021}
\setcopyright{acmcopyright}\acmConference[PEARC '21]{Practice and Experience in Advanced Research Computing}{July 18--22, 2021}{Boston, MA, USA}
\acmBooktitle{Practice and Experience in Advanced Research Computing (PEARC '21), July 18--22, 2021, Boston, MA, USA}
\acmPrice{15.00}
\acmDOI{10.1145/3437359.3465571}
\acmISBN{978-1-4503-8292-2/21/07}

\begin{document}

\SetKwFor{PFor}{parallel for}{do}{end}

\title{kEDM: A Performance-portable Implementation of Empirical Dynamic
Modeling using Kokkos}

\author{Keichi Takahashi}
\email{keichi@is.naist.jp}
\author{Wassapon Watanakeesuntorn}
\email{wassapon.watanakeesuntorn.wq0@is.naist.jp}
\author{Kohei Ichikawa}
\email{ichikawa@is.naist.jp}
\affiliation{%
  \institution{Nara Institute of Science and Technology}
  \city{Nara}
  \country{Japan}
}

\author{Joseph Park}
\email{josephpark@ieee.org}
\affiliation{%
  \institution{U.S. Department of the Interior}
  \city{Homestead}
  \state{Florida}
  \country{USA}
}

\author{Ryousei Takano}
\email{takano-ryousei@aist.go.jp}
\author{Jason Haga}
\email{jh.haga@aist.go.jp}
\affiliation{%
  \institution{National Institute of Advanced Industrial Science and Technology}
  \city{Tsukuba}
  \country{Japan}
}

\author{George Sugihara}
\email{gsugihara@ucsd.edu}
\affiliation{%
  \institution{University of California San Diego}
  \city{La Jolla}
  \state{California}
  \country{USA}
}

\author{Gerald M. Pao}
\email{pao@salk.edu}
\affiliation{%
  \institution{Salk Institute for Biological Studies}
  \city{La Jolla}
  \state{California}
  \country{USA}
}

\renewcommand{\shortauthors}{K. Takahashi et al.}

\begin{abstract}
    Empirical Dynamic Modeling (EDM) is a state-of-the-art non-linear time-series analysis framework. Despite its wide applicability, EDM was not
    scalable to large datasets due to its expensive computational cost. To
    overcome this obstacle, researchers have attempted and succeeded in accelerating EDM from
    both algorithmic and implementational aspects. In previous work, we
     developed a massively parallel implementation of EDM targeting
    HPC systems (mpEDM). However, mpEDM maintains different backends for different
    architectures. This design becomes a burden in the
    increasingly diversifying HPC systems, when porting to new hardware. In this paper, we design
    and develop a performance-portable implementation of EDM based on
    the Kokkos performance portability framework (kEDM), which runs on both CPUs and
    GPUs while based on a single codebase. Furthermore, we optimize individual kernels specifically for EDM computation, and use real-world datasets to demonstrate up to $5.5\times$ speedup compared to mpEDM
    in convergent cross mapping computation.
\end{abstract}



\maketitle

\section{Introduction}

\begin{figure*}
    \centering
    \includegraphics[width=\linewidth]{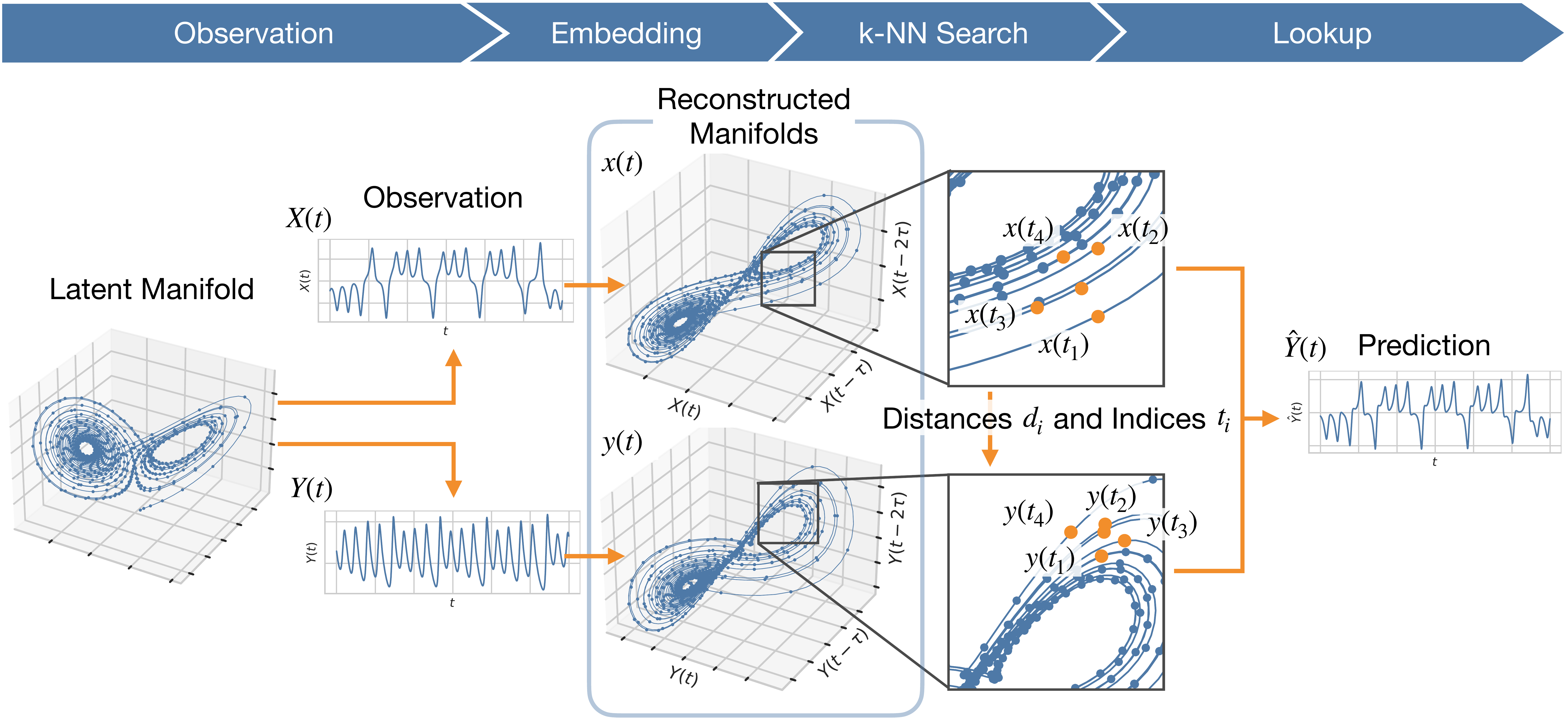}
    \caption{Overview of Convergent Cross Mapping (CCM)}\label{fig:edm}
\end{figure*}

Empirical Dynamic Modeling (EDM)~\cite{Chang2017} is a state-of-the-art
non-linear time-series analysis framework used for various tasks such as
assessing the non-linearity of a system, making short-term forecasts, and
identifying the existence and strength of causal relationships between
variables. Despite its wide applicability, EDM was not scalable to large
datasets due to its expensive computational cost. To overcome this challenge,
several studies have been conducted to accelerate EDM by improving the
algorithm~\cite{Ma2014} and taking advantage of parallel and distributed
computing~\cite{Pu2019}.

We tackle this challenge by taking advantage of modern HPC systems equipped
with multi-core CPUs and GPUs. We have been developing a massively
parallel distributed implementation of EDM optimized for GPU-centric HPC
systems, which we refer to as \textit{mpEDM}. In our previous work~\cite{mpedm}, we have deployed mpEDM on the AI
Bridging Cloud Infrastructure (ABCI)\footnote{\url{https://abci.ai}} to
obtain an all-to-all causal relationship map of all $10^5$ neurons in an
entire larval zebrafish brain. To date, this is the first causal analysis
of a whole vertebrate brain at single neuron resolution.

Although mpEDM has successfully enabled EDM computation at an unprecedented
scale, challenges remain. The primary challenge is
\textit{performance portability} across diverse hardware platforms. Recent HPC
landscape has seen significant increase in the diversity of processors and
accelerators. This is reflected in the design of upcoming exascale HPC systems, for example,
the Frontier system at the Oak Ridge National Laboratory will use AMD EPYC
CPUs and Radeon GPUs while the Aurora system at the Argonne National
Laboratory will employ Intel Xeon CPUs and Xe GPUs; additionally, the Fugaku system at RIKEN
uses Fujitsu A64FX ARM CPUs.

The current design of HPC applications has failed to keep up with this trend
of rapidly diversifying HPC systems. Computational application kernels are developed with the native programming model for the
respective hardware (\textit{e}.\textit{g}., CUDA on NVIDIA GPUs). mpEDM is no exception and
maintains two completely independent backends for x86\_64 CPUs and NVIDIA
GPUs. However, this design becomes a burden when supporting a diverse set of
hardware platforms because a new backend needs to be developed and maintained
for every platform. Based on this trend, various performance portability
frameworks~\cite{Deakin2019, Deakin2020} have emerged to develop
performance-portable applications from single codebase.

In this paper, we use the Kokkos~\cite{Edwards2014} framework and develop a performance-portable implementation of EDM\@ that runs on
both CPUs and GPUs, herein referred to as \textit{kEDM}. This new implementation is based on a single-source design and
facilitates the future development and porting to new hardware. Furthermore,
we identify and take advantage of optimization opportunities in mpEDM and
achieve up to 5.5$\times$ higher performance on NVIDIA V100 GPUs.

The rest of this paper is organized as follows: Section~\ref{sec:background}
first introduces EDM briefly and discusses the challenges in mpEDM\@;
Section~\ref{sec:proposal} presents kEDM, a novel implementation of EDM based on the Kokkos performance portability framework;
Section~\ref{sec:evaluation} compares kEDM and mpEDM using both synthetic and real-world datasets and assesses the efficiency of kEDM\@; and finally
Section~\ref{sec:conclusion} concludes the paper and discusses future work.

\section{Background}\label{sec:background}

\subsection{Empirical Dynamic Modeling (EDM)}\label{sec:edm}

Empirical Dynamic Modeling (EDM)~\cite{Chang2017,Ye2015} is a non-linear time
series analysis framework based on the Takens' embedding
theorem~\cite{Takens1981,Deyle2011}. Takens' theorem states that given a time-series
observation of a deterministic dynamical system, one can reconstruct the
latent attractor manifold of the dynamical system using time-delayed
embeddings of the observation. While the reconstructed manifold might not
preserve the global structure of the original manifold, it preserves the local
topological features (\textit{i}.\textit{e}., a diffeomorphism).

Convergent Cross Mapping (CCM)~\cite{Sugihara2012,Natsukawa2017,VanBerkel2020}
is one of the widely used EDM methods that identifies and quantifies the causal
relationship between two time series variables. Figure~\ref{fig:edm} illustrates the
overview of CCM\@. To assess if a time series $Y(t)$ (hereinafter called \textit{target})
causes another time seris $X(t)$ (hereinafter called \textit{library}), CCM performs
the following four steps:

\begin{enumerate}
    \item \textit{Embedding}: Both time series $X$ and $Y$ are embedded into
        $E$-dimensional state space using their time lags. For example, embedding
        of $X$ is denoted by $x$, where $x(t)=(X(t), X(t-\tau),
        \dots, X(t-(E-1) \tau))$. Here, $\tau$ is the time lag and $E$ is
        the embedding dimension, which is
        empirically determined and usually $E<20$ in real-world datasets.
    \item \textit{k-Nearest Neighbor Search}: For every library point $x(t)$,
         its $E+1$ nearest neighbors in the state space are searched. These
         neighbors form an $E$-dimensional simplex that encloses $x(t)$ in the
         state space. We refer to these nearest neighbors as $x(t_1)$, $x(t_2)$,
         \dots, $x(t_{E+1})$ and the Euclidean distance between $x(t)$ and
         $x(t_i)$ as $d(t, t_i) =\lVert x(t) - x(t_i)
        \rVert$.
    \item \textit{Lookup}: The prediction $\hat{y}(t)$ for a target point $y(t)$ is a
        linear combination of its neighbors $y(t_1), y(t_2), \dots, y(t_{E+1})$.
        Specifically,
        \begin{equation*}
            \hat{y}(t) = \sum^{E+1}_{i=1} \frac{w_i}{\sum^{E+1}_{i=1}{w_i}} \cdot y(t_i)
        \end{equation*}
        where
        \begin{equation*}
            w_i = \exp\left\{ -\frac{d(t, t_i)}{\min\limits_{1\leq i \leq E}{d(t, t_i)}}\right\}
        \end{equation*}
        The prediction $\hat{Y}(t)$ for $Y(t)$ is made by extracting the first
        component of $\hat{y}(t)$.
    \item \textit{Assessment of Prediction}: Pearson's correlation $\rho$
        between the target time series $Y$ and the predicted time series $\hat{Y}$
        is computed to assess the predictive skill.
        If $\rho$ is high, we conclude that $Y$ ``CCM-causes'' $X$.
\end{enumerate}

These four steps are repeated for all pair of time series when performing pairwise
CCM between multiple time series. Out of these steps, the $k$-nearest neighbor
search and the lookup consume significant runtime and need to optimized.
The $k$-nearest neighbor search and lookup in the state space are fundamental operations in EDM and generally dominate the runtime in other EDM algorithms as well.
We showed in our previous work~\cite{mpedm} that one can precompute the nearest
neighbors for every point in $x$ (all $k$-NN search) and store them as a lookup
table of distances and indices. This table can then be used to make
predictions for all target time series. This approach reduces the number of
$k$-NN search and provides significant speedup.

\subsection{Challenges in mpEDM}\label{sec:challenges}

This subsection discusses the two major challenges in
mpEDM\footnote{\url{https://github.com/keichi/mpEDM}}, our previous
parallel implementation of EDM\@.

\subsubsection{Performance portability across diverse hardware}\label{sec:portability}

The GPU backend of mpEDM was based on ArrayFire~\cite{Malcolm2012}, a general
purpose library for GPU computing. We chose ArrayFire primarily for its
productivity and portability. ArrayFire provides CUDA and OpenCL backends to
run on OpenCL devices. Although it also provides a CPU backend, most of the
functions provided by the CPU backend are neither multi-threaded nor
vectorized. We therefore developed our own implementation for CPUs
using OpenMP\@.

As a result, mpEDM had an ArrayFire-based GPU implementation and an OpenMP-based
CPU implementation of EDM, which double the maintenance cost. Considering the
diversifying target hardware, a unified implementation that achieves
consistent and reasonable performance across a diverse set of hardware is
required.

\subsubsection{Kernels tailored for EDM}\label{sec:flexibility}

Since mpEDM was relying on ArrayFire's k-nearest neighbor search function \texttt{nearestNeighbour()}, we were unable to modify or tune the k-NN
search to suit our use case and missed opportunities for further optimization.
ArrayFire's nearest neighbor function accepts arrays of reference and query
points, and returns arrays of closest reference points and their distances
for every query point. This interface is simple and easy-to-use; however, when
applying to EDM, the time-delayed embedding needs to be performed on the CPU in
advance and then passed on to ArrayFire. This hinders performance because it
increases the amount of memories copies between the CPU and the
GPU and memory reads from GPU memory.

Another potential optimization opportunity is the partial sort function
\texttt{topk()} invoked in the k-NN search. ArrayFire uses NVIDIA's
CUB\footnote{\url{https://nvlabs.github.io/cub}} library to implement partial
sort. CUB is a collection of highly optimized parallel primitives and is being
used by other popular libraries such
as Thrust\footnote{\url{https://github.com/NVIDIA/thrust}}. ArrayFire's
\texttt{topk()} function divides the input array into equal sized sub-array
and then calls CUB's parallel radix sort function to sort each sub-array. It
then extracts the top-$k$ elements from each sub-array and concatenates them
into a new array. This is recursively repeated until the global top-$k$
elements are found. Even though this implementation is well-optimized, it may
not be optimal for EDM because the target $k$ is relatively small ($\leq 20$).

Finally, we were unable to implement efficient lookups on GPU using ArrayFire.
ArrayFire provides a construct for embarrassingly parallel computation called
\textit{GFOR} that allows one to perform independent for-loops in parallel.
Although we were able to implement lookups using GFOR, the attained
performance was poor. Managing memory consumption and memory copies was also
challenging because ArrayFire implicitly allocates, deallocates and copies
arrays.

\section{kEDM}\label{sec:proposal}


\subsection{Overview}

kEDM\footnote{\url{https://github.com/keichi/kEDM}} is our
performance-portable implementation of EDM based on the Kokkos framework. We retain the high-level design of mpEDM, but
reimplement the whole application using Kokkos and optimize the bottleneck kernels
(\textit{i}.\textit{e}., all $k$-nearest neighbor search and lookup). To ensure the correctness of the
implementation, outputs from kEDM are validated against mpEDM as well as
the reference implementation of EDM
(cppEDM\footnote{\url{https://github.com/SugiharaLab/cppEDM}}), using automated unit tests.

Prior to implementing kEDM, we have examined a number of popular performance
portability frameworks. These include OpenMP, OpenACC, OpenCL and SYCL\@. We
chose Kokkos  because recent studies~\cite{Martineau2017, Deakin2019, Deakin2020}
have shown that it delivers portable performance on a large set of devices
compared to its alternatives. In addition, it has already been adopted by
multiple production applications~\cite{Sprague2020,Holmen2017,Demeshko2019}.
SYCL and OpenMP are certainly attractive choices as they have grown rapidly over the
last few years in terms of hardware coverage and delivered performance, but we
still consider them immature compared to Kokkos at the point of writing this
paper. Therefore, we choose Kokkos to implement kEDM\@.

\subsection{Kokkos}

Kokkos~\cite{Edwards2014} is a performance portability framework developed at
the Sandia National Laboratories. The aim of Kokkos is to abstract away the
differences between low-level programming models such as OpenMP, CUDA and HIP,
and exposes a high-level C++ programming interface to the developer. Kokkos allows
developers to build cross-platform and performance-portable applications on a
single codebase.

To parallelize a loop in the Kokkos programming model, the developer specifies (1)
the parallel pattern, (2) computational body, and (3) execution policy of the
loop. Available parallel patterns include parallel-for, parallel-reduce and
parallel-scan. The computational body of a loop is given as a lambda function.

The execution policy defines how a loop is executed. For example,
\texttt{RangePolicy} defines a simple 1D range of indices. \texttt{TeamPolicy}
and \texttt{TeamThreadRange} are used to launch teams of threads to exploit
the hierarchical parallelism of the hardware. For example, on a GPU, teams and threads map
to thread blocks and threads within thread blocks, respectively. On a CPU,
teams map to physical cores and threads map to hardware threads within cores.
\texttt{TeamPolicy} gives access to team-private and thread-private scratch
memory, an abstraction of shared memory in GPUs. Each execution policy is
bound to an execution space, an abstraction of where the code runs. The latest
release of Kokkos supports Serial, OpenMP, OpenMP Target, CUDA, HIP, Pthread,
HPX and SYCL as execution spaces.

\textit{Views} are fundamental data types in Kokkos that represent
homogeneous multidimensional arrays. Views are explicitly allocated by the
user and deallocated automatically by Kokkos using reference counting. Each
view is associated to a memory space, an abstraction of where the data
resides.

Listing~\ref{lst:basic} shows a vector addition  kernel implemented in Kokkos.
In this example, a parallel-for loop is launched that iterates over the 1D
range $[1,N]$.
Listing~\ref{lst:hierarchical} shows a matrix vector multiplication kernel
leveraging hierarchical parallelism. The outer parallel-for loop launches $M$ teams
that each computes one row of the output vector $y$. The inner parallel-reduce
computes the dot product between one row in $A$ and $x$.

\begin{lstlisting}[caption={Basic data parallel loop},label={lst:basic},float]
Kokkos::parallel_for(RangePolicy<ExecSpace>(N),
KOKKOS_LAMBDA(int i) {
    y(i) = a * x(i) + y(i);
});
\end{lstlisting}

\begin{lstlisting}[caption={Hierarchical data parallel loop},label={lst:hierarchical},float]
parallel_for(TeamPolicy<ExecSpace>(M, AUTO),
KOKKOS_LAMBDA(const member_type &member) {
    int i = member.team_rank();
    float sum = 0.0f;

    parallel_reduce(TeamThreadRange(member, N),
    [=] (int j, float &tmp) {
        tmp += A(i, j) * x(j)
    }, sum);

    single(PerTeam(member),
    [=] () {
        y(i) = sum;
    });
});
\end{lstlisting}

\subsection{All \texorpdfstring{$k$}{k}-Nearest Neighbor Search}

We implement the all $k$-NN search using an exhaustive approach similar
to~\cite{Garcia2010}. That is, we first calculate the pairwise
distances between all embedded library points in the state space and obtain a
pairwise distance matrix. Subsequently, each row of the obtained distance
matrix is partially sorted to find the distances and indices of the top-$k$
closest points.

\subsubsection{Pairwise distances}
As discussed in Section~\ref{sec:challenges}, storing the time-delayed
embeddings in memory and then calculating the pairwise distances is
inefficient since it increases memory access. Instead, we simultaneously
perform the time delayed embedding and distance calculation.

Algorithm~\ref{alg:distances} shows the pairwise distance calculation
algorithm in kEDM\@. Using Kokkos' \texttt{TeamPolicy} and
\texttt{TeamThreadRange}, we map the outer-most $i$-loop to thread teams and the $j$-loop to
threads within a team. A consideration on CPU is which loop to vectorize.
Since the inner-most $k$-loop is short ($E \leq 20$) in our use case,
vectorizing it is not profitable. We therefore use Kokkos'
\textit{SIMD types}\footnote{\url{https://github.com/Kokkos/simd-math}} to vectorize the
$j$-loop. SIMD types are short vector with overloaded operators that map to
intrinsic functions. SIMD types are mapped to scalars on GPUs and do not
impose any overhead. Note that the library time series $x$ is reused if $E > 1$
and we can expect more reuse with larger $E$. In addition, we explicitly cache
$x(k \tau + i)$ (where $k=[1, E]$) on team-local scratch
memory to speed up memory access.

\begin{algorithm}
    \SetAlgoLined
    \DontPrintSemicolon
    \KwIn{Library time series $x$ of length $L$}
    \KwOut{$L \times L$ pairwise distance matrix $D$}
    \tcp{\texttt{TeamPolicy}}
    \PFor{$i \leftarrow 1$ \KwTo $L$}{
        \tcp{\texttt{TeamThreadRange}}
        \PFor{$j \leftarrow 1$ \KwTo $L$}{
            $D(i, j) \leftarrow 0$\;
            \For{$k \leftarrow 1$ \KwTo $E$}{
                $D(i, j) \leftarrow D(i, j) + (x(k \tau + i) - x(k \tau + j))^2$\;
            }
        }
    }
    \caption{Pairwise distances}%
    \label{alg:distances}
\end{algorithm}

\subsubsection{Top-$k$ search}

The top-$k$ search kernel is particularly challenging to implement in a performance-portable manner
because state-of-the-art top-$k$ search algorithms~\cite{Johnson2019,Shanbhag2018}
are usually optimized for specific hardware. Thus, we designed and
implemented a top-$k$ search algorithm that works on both CPU and GPU
efficiently.

Algorithm~\ref{alg:partial-sort} outlines our top-$k$ search algorithm. In our algorithm, each
thread team finds the top-$k$ elements from one row of the distance matrix. Each thread within a thread team maintains a local priority queue on
team-private scratch memory that holds the top-$k$ elements it has seen so far. Threads read the distance matrix in a coalesced manner and push the distances and indices to their
local priority queues. Once all elements are processed, one leader thread in each thread team merges
the local queues within the team and writes the final top-$k$ elements to
global memory.

\begin{algorithm}
    \SetAlgoLined
    \DontPrintSemicolon
    \KwIn{$L \times L$ pairwise distance matrix $D$}
    \KwOut{$L \times k$ top-$k$ distance matrix $D_k$ and index matrix $I_k$}
    \tcp{\texttt{TeamPolicy}}
    \PFor{$i \leftarrow 1$ \KwTo $L$}{
        \tcp{\texttt{TeamThreadRange}}
        \PFor{$j \leftarrow 1$ \KwTo $L$}{
            Insert $D(i, j),\,j$ into local priority queue\;
        }
        \For{$j \leftarrow 1$ \KwTo $k$}{
            \For{$j \leftarrow 1$ \KwTo \# of threads in the team}{
                $D_k(i, j),\,I_k(i, j) \leftarrow$ Pop element from priority queue\;
            }
        }
        \tcp{Normalize $D_k$}
    }
    \caption{Partial sort}%
    \label{alg:partial-sort}
\end{algorithm}

\subsection{Lookup}

To increase the degree of parallelism and reuse of precomputed distance and
index matrices, we implement batched lookups. That is, we perform the lookups
for multiple target time series in parallel. CCM requires the library
time series to be embedded in the optimal embedding dimension of the target
time series. Therefore, we first group the target time series by their optimal
embedding dimensions and then perform multiple lookups for target time series that
have the same optimal embedding dimension in parallel.

Algorithm~\ref{alg:lookup} shows our lookup algorithm. The outer most $i$-loop
iterates over all time series of which optimal embedding dimension is $E$. The loop is
parallelized using \texttt{TeamPolicy}, where each team performs prediction of one
target time series. The $j$-loop is parallelized using \texttt{TeamThreadRange},
where each thread makes a prediction for each time point within a time series. The
inner most serial $k$-loop is unrolled to increase instruction-level parallelism. Since Kokkos currently
lacks a feature to indicate loop unrolling, we use the \texttt{\#pragma
unroll} directive here. The loop body requires indirect access to the target
time series using the indices table. To speed up random memory access, we cache the
target time series in team-private scratch memory.

\begin{algorithm}
    \SetAlgoLined
    \DontPrintSemicolon
    \KwIn{Array of target time series $y$, top-$k$ distance matrix $D_k$ and index
    matrix $I_k$ computed from the library time series}
    \KwOut{Array of predicted time series $\hat{y}$}
    \tcp{\texttt{TeamPolicy}}
    \PFor{$i \leftarrow 1$ \KwTo $N$}{
        \tcp{\texttt{TeamThreadRange}}
        \PFor{$j \leftarrow 1$ \KwTo $L$}{
                $\hat{y}(i, j) \leftarrow 0$\;
            \For{$k \leftarrow 1$ \KwTo $E+1$}{
                $\hat{y}(i, j) \leftarrow \hat{y}(i, j) + D_k(j, k) \cdot y(I_k(j, k))$ \;
            }
        }
    }
    \caption{Lookup}%
    \label{alg:lookup}
\end{algorithm}

In case the raw prediction is unneeded but only the assessment of the
predictive skill is needed, kEDM does not write out the predicted time series
to global memory. Instead, Pearson’s correlation is computed on-the-fly.
Kokkos’ custom reduction feature is used to implement parallel calculation of the
correlation coefficient. The algorithm is based on a numerically stable
algorithm for computing covariance proposed in~\cite{Schubert2018}.

\section{Evaluation}\label{sec:evaluation}

In this section, we compare kEDM with mpEDM using micro benchmarks and
real-world datasets. Furthermore, we conduct a roofline analysis of kEDM to
assess the efficiency of our implementation.

\subsection{Evaluation Environment}


We evaluated kEDM on two compute servers installed at the Salk Institute: Ika and Aori.
Ika is equipped with two sockets of 20-core Intel Xeon Gold
6148 CPUs, one NVIDIA V100 PCIe card and 384 GiB of DDR4 RAM\@. Aori is equipped with
two sockets of 64-core AMD EPYC 7742 CPUs and 1 TiB of DDR4 RAM\@. kEDM was built
with Kokkos 3.2 on both machines. On Aori, we used the AMD Optimizing C/C++
Compiler (AOCC) 2.2.0, a fork of Clang by AMD. On Ika, we used the NVIDIA CUDA
Compiler (NVCC) 10.1.

\subsection{Micro Benchmarks}

We implemented micro benchmarks to measure the performance of the individual
kernels we described in Section~\ref{sec:proposal} and compare it with that of
mpEDM\@.

\begin{figure}
    \centering
    \includegraphics[width=.80\linewidth]{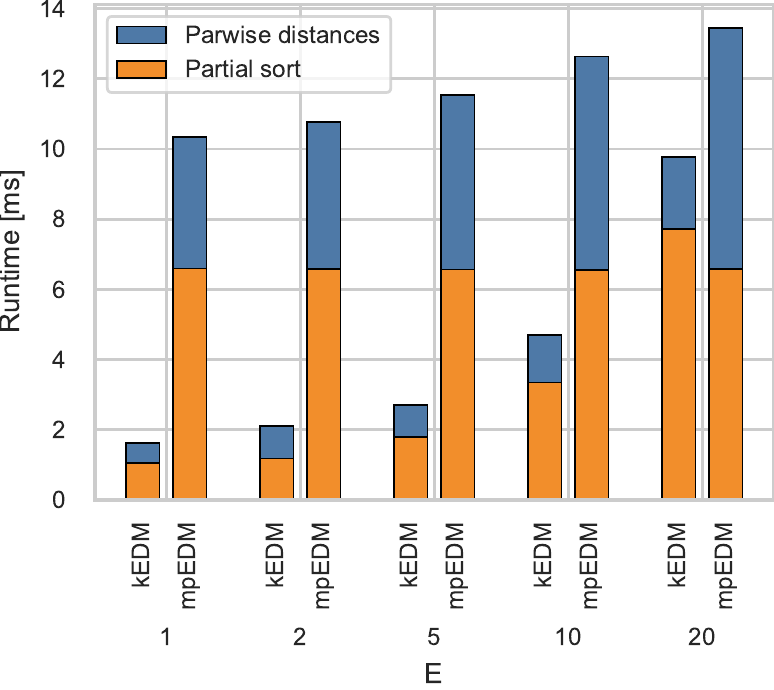}
    \caption{Breakdown of all k-NN search runtime on V100 ($L=10^4$)}%
    \label{fig:breakdown-knn-v100}
\end{figure}

\begin{figure}
    \centering
    \includegraphics[width=.80\linewidth]{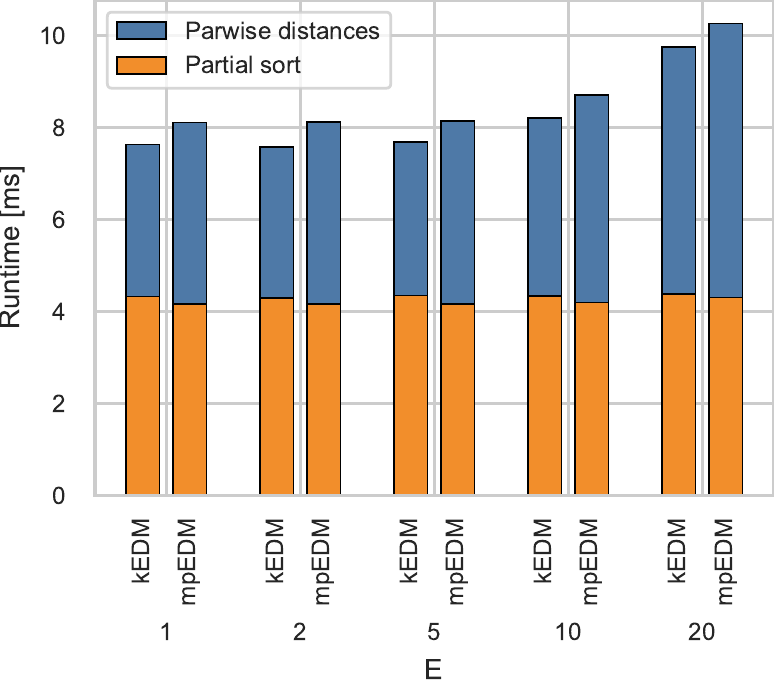}
    \caption{Breakdown of all k-NN search runtime on EPYC 7742 ($L=10^4$)}%
    \label{fig:breakdown-knn-epyc}
\end{figure}

Figure~\ref{fig:breakdown-knn-v100} shows the runtime of the all k-NN search kernel
of kEDM and mpEDM on NVIDIA V100. Here, we generated a synthetic time series with
$10^4$ time steps and varied the embedding dimension.
The results indicate that the pairwise
distance calculation in kEDM is significantly faster than mpEDM (up to
6.6$\times$). This is because the time-delayed embedding is performed during the distance
calculation on the GPU\@. The partial sort is also faster than mpEDM if $E$ is small
(6.2$\times$ faster if $E=1$). However, kEDM's partial sort performance degrades as
$E$ increases and  slightly underperforms mpEDM when $E=20$. Since the local
priority queues are stored in shared memory, increasing the capacity of the local
queues increases shared memory usage and results in lower multiprocessor
occupancy. We confirmed this fact using NVIDIA's Nsight Compute profiler.
Figure~\ref{fig:breakdown-knn-epyc} shows the runtime on AMD EPYC 7742. kEDM
exhibits identical performance as mpEDM on EPYC 7742.

\begin{figure}
    \centering
    \includegraphics[width=.80\linewidth]{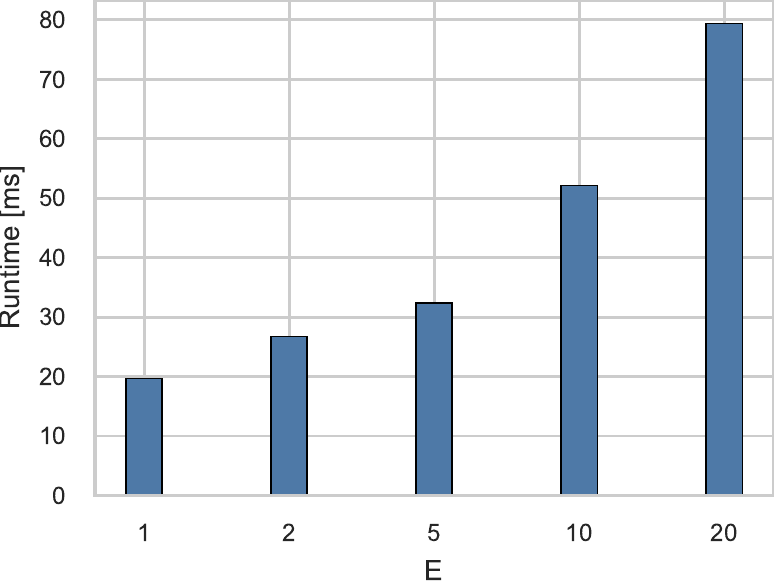}
    \caption{Runtime of lookups on V100 ($L=10^4, N=10^5$)}%
    \label{fig:breakdown-lookup-v100}
\end{figure}

\begin{figure}
    \centering
    \includegraphics[width=.80\linewidth]{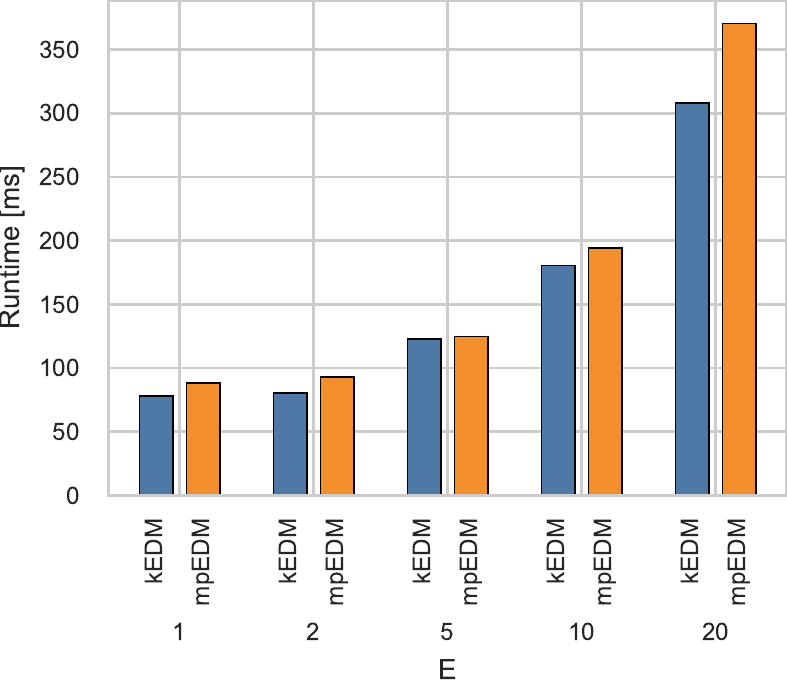}
    \caption{Runtime of lookups on EPYC 7742 ($L=10^4, N=10^5$)}%
    \label{fig:breakdown-lookup-epyc}
\end{figure}

Figures~\ref{fig:breakdown-lookup-v100} and~\ref{fig:breakdown-lookup-epyc}
show the runtime of the lookup kernel (without cross correlation calculation) on V100 and EPYC 7742, respectively.
Here, we generated $10^5$ synthetic target time series each having $10^4$ time
steps and performed lookups from a library time series with the same length.
Note that we only executed kEDM on
V100 since mpEDM lacks a lookup kernel for GPU as described earlier in
Section~\ref{sec:flexibility}. These plots indicate that kEDM on V100
consistently outperforms EPYC 7742 by a factor of 3--4$\times$.
Interestingly, kEDM slightly outperforms mpEDM on EPYC as well.
This might attribute to the fact that we load the target time series to scratch memory before performing lookups. Even though CPUs do not have user-manageable memory as opposed to GPUs, accessing the
target time series might have loaded the time series on cache and improved the
cache hit ratio.

\subsection{Real-world Datasets}

We prepared six real-world datasets with diverse number and length of time
series that reflect the variety of use cases. We then measured the runtime of kEDN for completing pairwise CCM calculations.

Table~\ref{tbl:real-world} shows the runtime for performing a pairwise CCM on
each dataset.
Fish1\_Normo is a subset of 154 representative neurons of the dominant default zebrafish larvae neuronal behaviors collected by lightsheet microscopy of fish transgenic with nuclear localized GCAMP6f, a calcium indicator. Fly80XY is a drosophila melanogaster whole brain lightfield microscopy GCAMP6 recording, where distinct brain areas were identified by independent component analysis with the fly left right and forward walking speed behaviors collected on a styrofoam ball. Genes\_MEF contains the gene expression profiles of all genes and small RNAs from mouse embryo fibrobast genes over 96 time steps of two cycles of serum induction and starvation stimulation. Subject6 and Subject11 are whole brain light sheep microscopy GCAMP6f recordings at whole brain scale and single neuron resolution of larval zebrafish. F1 is a subset of a larval zebrafish biochemically induced seizure recording with three phases: control conditions, pre-seizure and full seizure.

The results clearly demonstrate that kEDM outperforms mpEDM in most
cases. In particular, kEDM shows significantly higher (up to 5.5$\times$)
performance than mpEDM on NVIDIA Tesla V100 and Intel Xeon Gold 6148.
This performance gain is obtained from the optimized $k$-nearest neighbor search and GPU-enabled lookup.

\begin{table*}
\centering
\caption{Benchmark of CCM runtime using real-world datasets}%
\label{tbl:real-world}
\begin{tabular}{@{}lrrrrrrrr@{}}
\toprule
             & \multicolumn{1}{c}{} & \multicolumn{1}{c}{} & \multicolumn{3}{c}{V100 \& Xeon Gold 6148 } & \multicolumn{3}{c}{EPYC 7742} \\ \cmidrule(l){4-6}  \cmidrule(l){7-9}
Dataset      & \multicolumn{1}{c}{\# of Time Series} & \multicolumn{1}{c}{\# of Time Steps} & kEDM & mpEDM & Speedup & kEDM & mpEDM & Speedup \\ \midrule
Fish1\_Normo &  154    & 1,600  &      3s &     11s & 3.67$\times$ &      3s &      4s & 1.33$\times$ \\
Fly80XY      &  82     & 10,608 &     11s &     50s & 4.55$\times$ &     22s &     30s & 1.36$\times$ \\
Genes\_MEF   &  45,318 & 96     &    344s &    334s & 0.97$\times$ &     96s &    139s & 1.45$\times$ \\
Subject6     &  92,538 & 3,780  &  5,391s & 29,579s & 5.49$\times$ & 12,145s & 11,571s & 0.95$\times$ \\
Subject11    & 101,729 & 8,528  & 20,517s & 85,217s & 4.15$\times$ & 43,812s & 38,542s & 0.88$\times$ \\
F1           &  8,520  & 29,484 & 11,372s & 20,128s & 1.77$\times$ & 23,001s & 19,950s & 0.87$\times$ \\ \bottomrule
\end{tabular}
\end{table*}

\subsection{Efficiency}

To assess the efficiency of our implementation, we conducted a roofline
analysis~\cite{Williams2008} of our kernels. The compute and memory ceilings
on each platform were measured using the Empirical Roofline Toolkit (ERT)\footnote{\url{https://bitbucket.org/berkeleylab/cs-roofline-toolkit}} 1.1.0.
Since ERT fails to measure the L1 cache bandwidth on GPUs, we used the
theoretical peak performance instead. We followed the methodology presented
in~\cite{Yang2020b} to measure the arithmetic intensity and the
attained FLOP/s. Nvprof 10.1 and likwid~\cite{Treibig2010} 5.0.1 were used to
collect the required metrics on GPU and CPU, respectively.
We used an artificial dataset containing $10^5$ time series each having $10^4$
time steps. This scale reflects our largest use case.

\begin{figure}
    \centering
    \includegraphics[width=.90\linewidth]{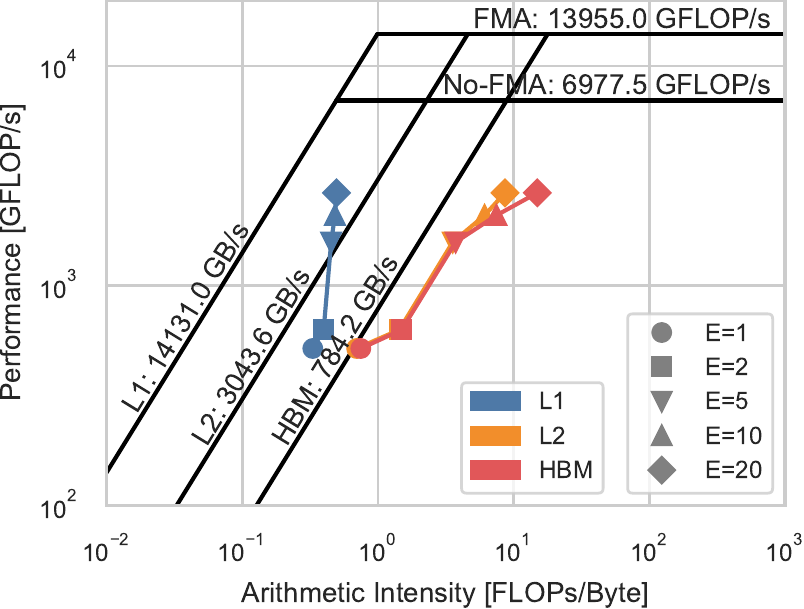}
    \caption{Roofline analysis of pairwise distance kernel on V100 ($L=10^4$)}%
    \label{fig:roofline-distances-v100}
\end{figure}
\begin{figure}
    \centering
    \includegraphics[width=.90\linewidth]{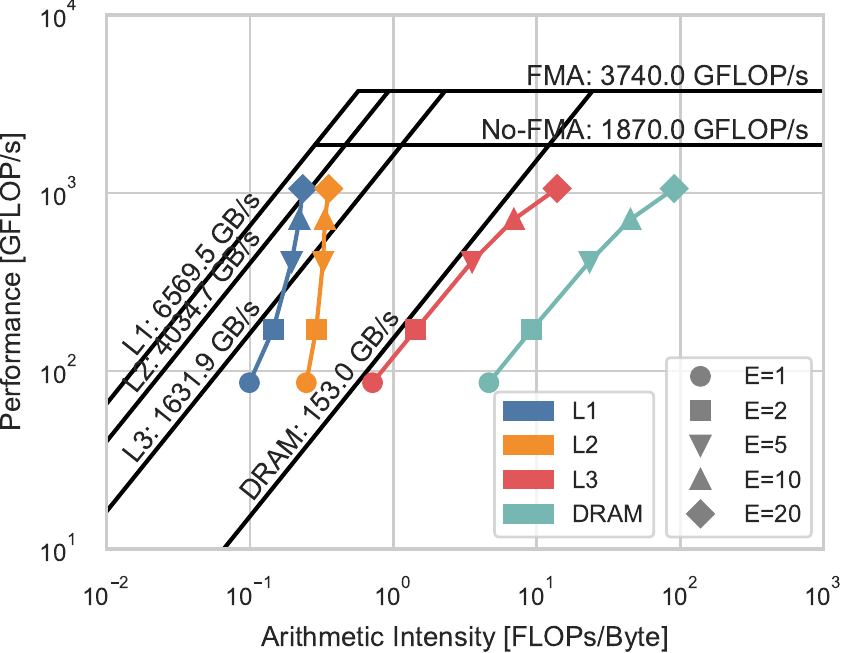}
    \caption{Roofline analysis of pairwise distance kernel on EPYC 7742 ($L=10^4$)}%
    \label{fig:roofline-distances-epyc}
\end{figure}

Figures \ref{fig:roofline-distances-v100} and \ref{fig:roofline-distances-epyc}
depict the hierarchical roofline models for the pairwise distance kernel on V100
and EPYC 7742, respectively. As expected, the arithmetic intensity of the
pairwise distance kernel increases with the embedding dimension since the
reuse of the time series improves. On V100, the kernel was bounded by HBM
bandwidth when $E=1$. However, the kernel was not able to reach the rooflines
as $E$ increases. We found out using NVIDIA Nsight profiler that the load/store
units were saturated because of excessive memory transactions. A common
technique to reduce the number of memory transactions is to use vectorized
loads and stores; however, it is not applicable here because the memory
alignment depends on the user-supplied parameters $E$ and $\tau$. On EPYC 7742,
the kernel is initially hitting the L3 cache roofline and then bounded by
L1 and L2 cache bandwidth.

\begin{figure}
    \centering
    \includegraphics[width=.90\linewidth]{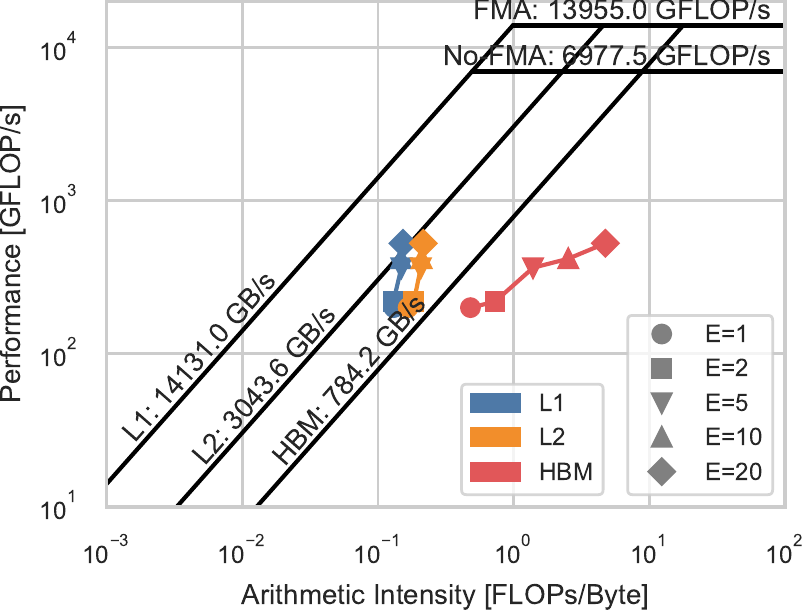}
    \caption{Roofline analysis of lookup kernel on V100 ($L=10^4, N=10^5$)}%
    \label{fig:roofline-lookup-v100}
\end{figure}

\begin{figure}
    \centering
    \includegraphics[width=.90\linewidth]{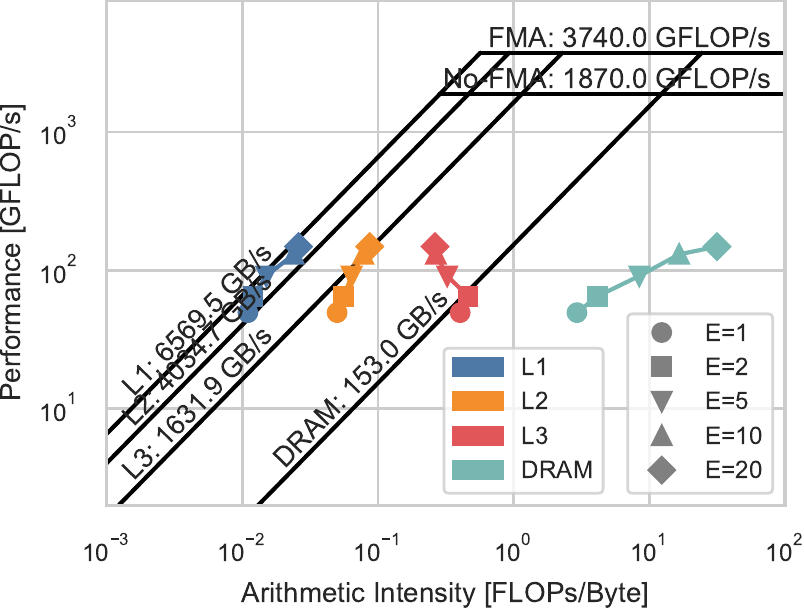}
    \caption{Roofline analysis of lookup kernel on EPYC 7742 ($L=10^4, N=10^5$)}%
    \label{fig:roofline-lookup-eypc}
\end{figure}

Figures \ref{fig:roofline-lookup-v100} and \ref{fig:roofline-lookup-eypc} present
the rooflines for the lookup kernel on V100 and EPYC 7742, respectively. These
plots suggest that the lookup kernel is bounded by the L2 cache on V100 and
the L1 cache on EPYC 7742. This is explained from the fact that the distance
and index matrices fit on the respective caches. For example, the total size of
the distance and index matrices is 1.6 MB if $E=20$, which fits on EPYC 7742's L1
cache (4 MiB) and V100's L2 cache (6 MiB).

Overall, these roofline models indicate that kEDM operates close to the
ceilings and achieves high efficiency in most cases. Furthermore, the roofline
models reveal that EDM is an inherently memory-bound algorithm, primarily
bounded by memory or cache bandwidth depending on the embedding dimension. It
does not enter the compute-bound region of the roofline model in our use cases.
This observation suggests that kEDM would benefit from hardware with higher memory,
cache, and load/store unit bandwidth.

\section{Conclusion \& Future Work}\label{sec:conclusion}

We designed and developed kEDM, a performance portable implementation of EDM
using the Kokkos performance portability framework. The new implementation
is based on a single codebase and runs on both CPUs and GPUs. Furthermore, we removed several
several inefficiencies from mpEDM and custom-tailored kernels. Benchmarks using real-world datasets indicate up to $5.5\times$
speedup in convergent cross mapping computation.

In the future, we will implement a Python binding to facilitate adoption by
users. We are also planning to evaluate on other hardware platforms such as AMD GPUs and Fujitsu A64FX ARM processors.

\begin{acks}
This work was partly supported by JSPS KAKENHI Grant Number JP20K19808 (KT) and an
Innovation grant by the Kavli Institute for Brain and Mind (GMP). The authors would like to thank Dominic R. W. Burrows at the MRC Centre for Neurodevelopmental Disorders, Institute of Psychiatry, Psychology and Neuroscience, King's College London, London, UK for providing the F1 dataset used in the performance evaluation.
\end{acks}

\bibliographystyle{ACM-Reference-Format}
\bibliography{references}

\end{document}